\documentclass{IEEEtran4PSCC}

\usepackage{nomencl}
\makenomenclature
\usepackage{siunitx}

\usepackage[caption=false,font=footnotesize]{subfig}	
\usepackage{xcolor}

\definecolor{blue}{RGB}{31, 119, 180}    
\definecolor{orange}{RGB}{255, 127, 14}  
\definecolor{green}{RGB}{44, 160, 44}    
\definecolor{red}{RGB}{214, 39, 40}      
\definecolor{purple}{RGB}{148, 103, 189} 

\usepackage{algorithm}
\usepackage{algorithmicx}
\usepackage{algpseudocode}
\usepackage{courier}

\usepackage{amsmath,amsfonts,amssymb,amsthm}
\usepackage{mathtools}
\usepackage{mathrsfs}	
\usepackage{subdepth}	

\usepackage{array}	
\usepackage[inline]{enumitem} 

\usepackage{hyperref}
\usepackage[sort,compress,capitalise]{cleveref}	
\usepackage{url}

\usepackage{xcolor} 
\usepackage{xspace} 
\usepackage{lipsum}

\usepackage{tikz}
\usetikzlibrary{babel}  
\usepackage[straightvoltages,nooldvoltagedirection]{circuitikz}
\usetikzlibrary{shapes,arrows}
\usepackage{verbatim}
\usepackage{multirow}

 \usepackage{booktabs}

\DeclareSIUnit{\rpm}{rpm}
\DeclareSIUnit{\var}{VAr}

{
\theoremstyle{plain}

}

{
\theoremstyle{definition}

}

\crefname{Hypothesis}{Hyp.}{Hyps.}
\Crefname{Hypothesis}{Hyp.}{Hyps.}

\crefname{Lemma}{Lemma}{Lemmata}
\Crefname{Lemma}{Lemma}{Lemmata}

\crefname{Definition}{Def.}{Defs.}
\Crefname{Definition}{Def.}{Defs.}

\newcommand{\DFT}[1][]{DFT\xspace}  

\usepackage{bm}
\usepackage{mathtools}

\ifCLASSINFOpdf
\else
   \usepackage[dvips]{graphicx}
\fi
\makeatletter
\let\old@ps@headings\ps@headings
\let\old@ps@IEEEtitlepagestyle\ps@IEEEtitlepagestyle
\def\psccfooter#1{%
    \def\ps@headings{%
        \old@ps@headings%
        \def\@oddfoot{\strut\hfill#1\hfill\strut}%
        \def\@evenfoot{\strut\hfill#1\hfill\strut}%
    }%
    \def\ps@IEEEtitlepagestyle{%
        \old@ps@IEEEtitlepagestyle%
        \def\@oddfoot{\strut\hfill#1\hfill\strut}%
        \def\@evenfoot{\strut\hfill#1\hfill\strut}%
    }%
    \ps@headings%
}
\makeatother

\psccfooter{%
        \parbox{\textwidth}{\hrulefill \\ \small{23rd Power Systems Computation Conference} \hfill \begin{minipage}{0.2\textwidth}\centering \vspace*{4pt} \includegraphics[scale=0.06]{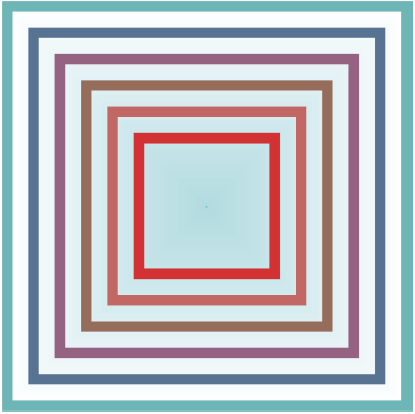}\\\small{PSCC 2024} \end{minipage} \hfill \small{Paris, France --- June 4 -- 7, 2024}}%
}

\begin{document}
%
\title{Enhanced Frequency Containment Reserve Provision from Battery Hybridized Hydropower Plants: Theory and Experimental Validation}


\author{\IEEEauthorblockN{Francesco Gerini\IEEEauthorrefmark{2},
Elena Vagnoni\IEEEauthorrefmark{1}, 
Martin Seydoux\IEEEauthorrefmark{1},
Rachid Cherkaoui\IEEEauthorrefmark{2} and
Mario Paolone\IEEEauthorrefmark{2}}
\IEEEauthorblockA{\IEEEauthorrefmark{2} Distributed Electrical System Laboratory (DESL) \\
EPFL,
Lausanne, Switzerland\\ }
\IEEEauthorblockA{\IEEEauthorrefmark{1} Technology Platform for Hydraulic Machines (PTMH)\\
EPFL,
Lausanne, Switzerland\\ }
}

\maketitle

\begin{abstract}
This paper presents a solution to address wear and tear of \emph{Run-of-River} (RoR) \emph{Hydropower Plants} (HPPs) providing enhanced \emph{Frequency Containment Reserve} (FCR). In this respect, the study proposes the integration of a \emph{Battery Energy Storage System} (BESS) with RoR HPPs controlled by a double-layer \emph{Model Predictive Control} (MPC). The upper layer MPC acts as a state of energy manager for the BESS, employing a forecast of the required regulating energy for the next hour. The lower-layer MPC optimally allocates the power set-point between the turbine and the BESS. Reduced-scale experiments are performed on a one-of-a-kind testing platform to validate the proposed MPC-based control considering a comparison with different control strategies and different BESS sizes. The results demonstrate superior performance of the proposed framework, compared to simpler techniques like dead-band control or to the standalone RoR scenario, leading to improved FCR provision, reduced servomechanism stress, and extended hydropower asset lifespan.
\end{abstract}


\thanksto{\noindent This work is funded by the XFLEX HYDRO project. The XFLEX HYDRO project has received funding from the European Union’s Horizon 2020 research and innovation programme under grant agreement No 857832. }
\section{Introduction}
\label{sec:Introduction}
As widely recognized, hydropower plants are renewable energy assets that play a crucial role in providing fundamental ancillary grid services, such as \emph{Frequency Containment Reserve} (FCR), which have become increasingly important due to the decommissioning of dispatchable thermal power plants. Part of the FCR reserve is provided by \emph{Run-of-River} (RoR) power plants \cite{hase_balancing_2021}, accounting for 5.94 \% of the total generated electricity in the ENTSOE area in 2022 \cite{entsoe_statistical_2022}. 

The need for continuous power regulations impacts the lifetime of hydroelectric assets \cite{yang_burden_2018}. RoR \emph{Hydropower Plants} (HPP) are often equipped with double regulated Kaplan turbines (i.e. machines able to control guide vanes and blades opening angles), able to guarantee high efficiency values through a wide range of water discharge and head conditions. In the case of Kaplan turbines, the lifetime of the servomechanism that controls the movement of the blades is significantly impacted by continuous regulation. Continuous movements can stress the servomechanisms, leading to increased wear and tear, potential mechanical failures, and reduced overall turbine life \cite{valentin_hybridization_2022}. To address these challenges, this paper focuses on the hybridization of RoR HPP with \emph{Battery Energy Storage Systems} (BESS) to enhance FCR provision and extend hydropower asset lifetime. 
This approach has been gathering an increasing interest in the literature \cite{makinen_modelling_2020,valentin_benefits_2022,kadam_hybridization_2023,cassano_model_2022}. However, despite this interest, many of the existing contributions primarily suggest simple control techniques based on dead-band control or fuzzy logic \cite{kadam_hybridization_2023, makinen_modelling_2020}. 
Others discuss HPP-BESS hybridization for other applications, such as penstock fatigue reduction in medium-head HPPs \cite{cassano_model_2022}. 
Moreover, most of the above-mentioned contributions are only simulation-based or with very limited experimental validation \cite{valentin_benefits_2022}). 

For this reason, the scope of this paper is to propose and experimentally validate an optimal control technique for hybrid RoR HPPs operating under a daily dispatch plan that provide FCR with a fixed droop characteristic. In particular, we present a double-layer \emph{Model Predictive Control} (MPC) to drive the hybrid system. The upper layer MPC (slower and farsighted) ensures the continuous operation of the BESS, by acting as \emph{State of Energy} (SOE) manager, leveraging a forecast of the regulating energy necessary to provide the FCR service in the following hour. The lower layer (faster and short-sighted) is responsible for splitting the requested power set-point between the turbine and the BESS, ensuring the feasible operation of both systems. The framework is validated for different BESS power and energy ratings to study the impact of the BESS sizing on the control problem. Furthermore, the efficacy of the proposed control strategy is examined and validated through reduced-scale experiments conducted on an innovative testing platform \cite{gerini_experimental_2021}. The evaluation covers comparison with classical control strategies and BESS sizes, with a specific emphasis on assessing the FCR provision quality and the reduction in servomechanism stress.

The paper is organized as follows. Section \ref{sec:Problem Statement} proposes the general formulation of the control problem. Section \ref{sec:Control Framework} presents a detailed description of the two-stage control framework. Section \ref{sec:Experimental Validation} provides the experimental validation of the proposed framework. Finally, Section \ref{sec:Conclusion} summarizes the original contributions and main results of the article and proposes perspectives for further research activities.

\section{Problem Statement}
\label{sec:Problem Statement}
As stated in \cref{sec:Introduction}, the control addresses run-of-river HPPs operating under daily dispatch plans $P^\text{disp}$, scheduled hourly, and obligated to provide FCR service with a fixed droop characteristic\footnote{Numerous European Transmission System Operators (TSOs) source their FCR in a shared market with 4-hour adjustments \cite{entsoe_tsos_2018}. However, HPP droop settings tend to remain fixed for extended periods (years or even decades) due to contractual agreements, making them constant inputs in our study.} $\sigma_f$. The dispatch plan, input to the problem, is the product of an external optimization, taking into account market prices and constraints such as the concession head limit and other variables, and not object of this study. As already stated, a BESS is integrated into the system. The primary focus of this study is to propose an optimal set-point splitting policy that effectively achieves the following objectives:
\begin{enumerate}[label=\roman*)]
    \item Ensuring dispatch tracking and high-quality FCR provision characterized by a rapid response time, in compliance with stringent FCR requirements \cite{entsoe_all_2019,swissgrid_ltd_conditions_2009};
    
    \item Minimizing the number of movements and the mileage of the hydropower servomechanisms;
    
    \item Ensuring the continuous and efficient operation of the BESS by managing its SOE within physically feasible limits;
    
    \item Validating the feasibility of the BESS power set-point to uphold operational constraints composed by the power converter capability curve.
\end{enumerate}
\begin{figure}[t] 
  \centering       
  \includegraphics[width=0.85\linewidth]{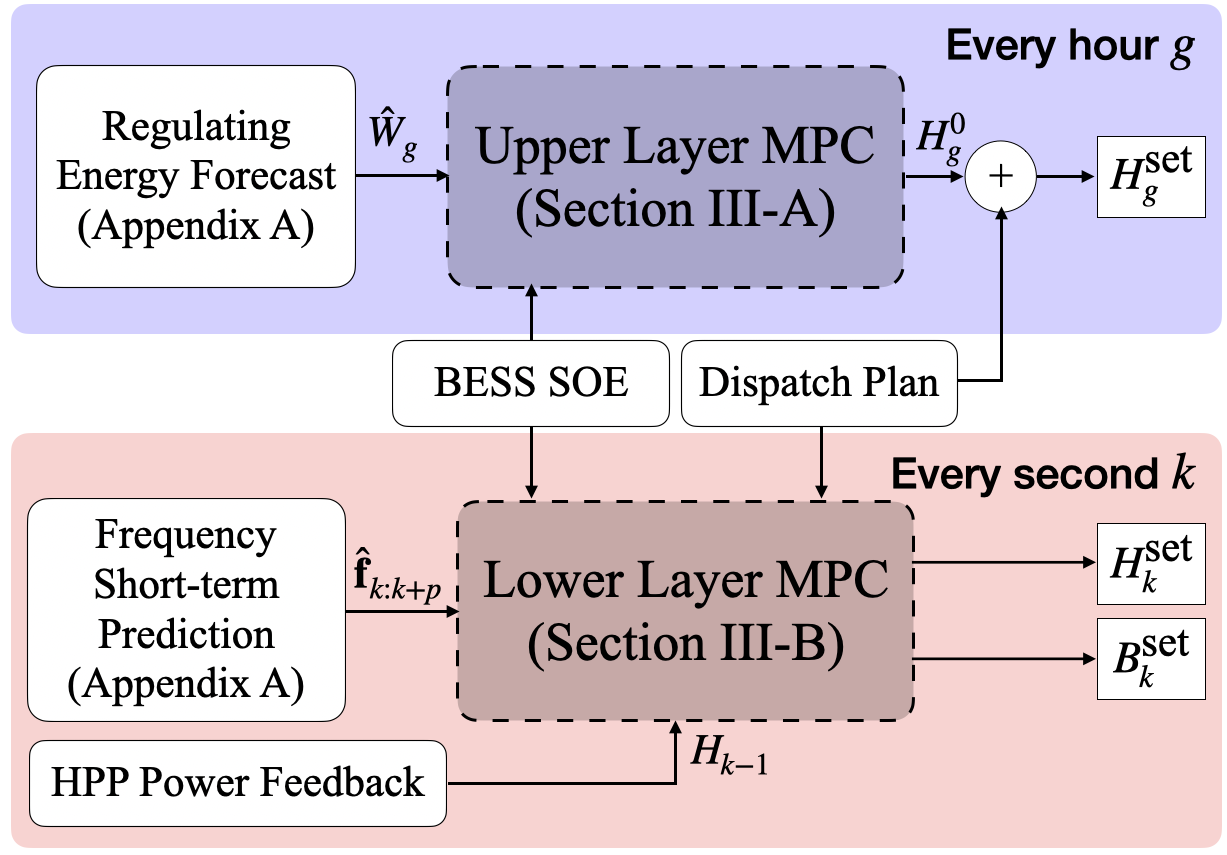} 
  \caption{DLMPC structure, showing inputs and output of the UL (in blue) and LL (in red).}
  \label{fig:Framework} 
\end{figure}
\begin{figure}[t] 
  \centering       
  \includegraphics[width=0.8\linewidth]{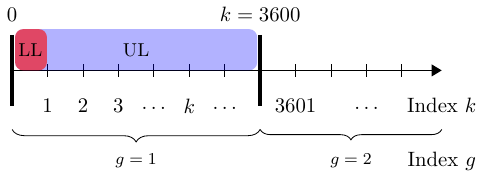} 
  \caption{Timeline of the problem, showing the actuation of the UL in blue, i.e., each hour $g$ and of the LL in red, i.e., each second $k$}
  \label{fig:timeline} 
\end{figure}
To achieve these objectives, the study proposes the use of a \emph{Double-Layer Model Predictive Control} (DLMPC), structured as visible in \cref{fig:Framework}. The \emph{ upper layer} (UL) acts as a SOE manager for the BESS, employing a forecast of the required regulating energy\footnote{The idea of forecasting the FCR regulating energy was first introduced by \cite{piero_schiapparelli_quantification_2018} and later used from \cite{gerini_improving_2021} and others. More information about the way this forecast is performed can be found in Appendix \ref{Appendix:FrequencyForecasting}.} for each hour $g$. The output of the UL is a constant power offset for the hour that directly modifies the dispatch plan of the HPP. The \emph{Lower Layer} (LL) optimally allocates the power set point between the turbine and the BESS in real time, that is, every second $k$ (see the time indices used in \cref{fig:timeline}). This control layer leverages short-term frequency prediction\footnote{Appendix \ref{Appendix:FrequencyForecasting} examines the viability of this forecast by making references to relevant literature.}, together with real time measurement of the HPP power and the BESS SOE. In \cref{fig:Framework}, each quantity $A$ is indicated with a certain time subscript: hour $g$ or seconds $k$. In particular, if the quantity is a vector, including information from time $k$ to $k+p$, it is indicated as:
\begin{equation}
    \mathbf{A}_{k:k+p} = \left[A_{k},A_{k+1}, A_{k+2},\ldots, A_{k+p-1},A_{k+p}\right] \nonumber.
\end{equation}
\section{Control Framework}
\label{sec:Control Framework}
\subsection{Upper Layer MPC}
\label{sec:UL}
Let us consider $H$ as the power of the hydroelectric unit (active sign notation) and $B$ as the power of BESS (passive sign notation).
This MPC layer, executed at the beginning of every hour $g$, is responsible for computing the smallest constant hourly power offset:
\begin{equation}
    H_g^0 = B_g^0 \nonumber
\end{equation}
that allows to keep the BESS SOE within its physical limits.  
The problem relies on the last measure of the BESS SOE and on a forecast of the frequency integral over the next hour $\hat{W}_g$, together with its confidence intervals $\hat{W}_g^{\uparrow}, \hat{W}_g^{\downarrow}$. Information on the forecasting method can be found in Appendix \ref{Appendix:FrequencyForecasting}. The forecast is modified to take into account the charging efficiency $\eta_{\text{ch}}$ and the discharging efficiency $\eta_{\text{dch}}$ of the BESS as:
\begin{equation}
 W_g  = \left\{\begin{array}{cc}
         \hat{W}_g\cdot \eta_{\text{ch}} , & \text{if } \hat{W}_g\le 0  \\
         \hat{W}_g / \eta_{\text{dch}} , & \text{if } \hat{W}_g\ge 0  \\
    \end{array}  \right.  \label{eq:Wh} \\
\end{equation}
The updated value $W_g$ is an input for the upper layer MPC, where it is used together with the frequency droop $\sigma_f$ to forecast the regulating energy for hour $g$ for the FCR service:
\begin{equation}
    E_g^{\text{FCR}} =\sigma_f \cdot W_g
\end{equation}
The constant hourly offset $B_g^0$ is either positive (BESS is charging) or negative (BESS is discharging):
\begin{equation}
B_g^0  = \left\{\begin{array}{cc}
         \phantom{-}B_g^{0+} , & \text{if } B_g^0 \ge 0, \\
         -B_g^{0-} , & \text{if } B_g^0 \le 0
    \end{array}  \right.  \label{eq:B_g} \\
\end{equation} where  $B_g^{0+}$ and $B_g^{0-}$ are the charging and discharging BESS power in $\SI{}{\kilo\watt}$, respectively.
The constant hourly offset $B_g^0$ is then computed according to the following \emph{Optimisation Problem} (OP):
\begin{subequations}
\begin{alignat}{1}
        B_g^0 & = \underset{B_g^{0+} \& B_g^{0-} \in \mathbb{R}}{\arg \min } \left(B_g^{0+}+B_g^{0-}\right)^2 \label{eq:UL_Objective} \\
         \text { s.t. } 
        &  W_g  = \left\{\begin{array}{cc}
         \hat{W}_g\cdot \eta_{\text{ch}} , & \text{if } \hat{W}_g\le 0  \\
         \hat{W}_g / \eta_{\text{dch}} , & \text{if } \hat{W}_g\ge 0  \\
    \end{array}\right.  \label{eq:UL_Wg_forecast}\\
        &     E_g^{\text{FCR}} =\sigma_f \cdot W_g \label{eq:UL_RegulatingEnergy}\\
        & E_g^0 = \eta_{\text{ch}}\cdot B_g^{0+}-\frac{1}{\eta_{\text{dch}}} \cdot B_g^{0-} \label{eq:UL_SOE_OFFSET}\\
        & SOE_{g}=SOE_{h-1} + \left[ E_{g}^0 +E_g^{\text{FCR}}\right] \frac{1}{C_{\mathrm{B}}} \label{eq:UL_SOE} \\
        & {SOE}_{g}^{\uparrow}=SOE_{g}+\left[\sigma_f \cdot \hat{W}_g^{\uparrow}\right] \frac{1}{C_{\mathrm{B}}}  \label{eq:UL_SOE_up}\\
        & {SOE}_{g}^{\downarrow}={SOE}_{g}-\left[\sigma_f \cdot \hat{W}_g^{\downarrow}\right] \frac{1}{C_{\mathrm{B}}} \label{eq:UL_SOE_down}\\
        & {SOE}_{g}^{\uparrow} \leq SOE_{\max} \label{eq:UL_SOE_up_LIM}\\
        & {SOE}_{g}^{\downarrow} \geq SOE_{\min} \label{eq:UL_SOE_down_LIM}\\
        &0 \leq B_g^{0+} \leq B_{\max }^{+}\label{eq:UL_limitCharging}\\
        & 0 \leq B_g^{0-} \leq B_{\max }^{-} \label{eq:UL_limitDischarging}\\
        & B_g^0=B_g^{+}-B_g^{-} \label{eq:UL_BESSPower}
    \end{alignat}
    \label{eq:UpperLayerMPC}
\end{subequations} The energy variation $E_g^0$ due to the offset action during hour $g$ is provided by \cref{eq:UL_SOE_OFFSET} and expressed in $\SI{}{\kilo\watt\hour}$, as the power offset is applied for one hour. The latter is used, together with the forecast of $W_g$ in \cref{eq:UL_SOE} to predict the $SOE_g$ at the end of hour $g$, taking into consideration the BESS capacity $C_{\text{B}}$. Eqs. (\ref{eq:UL_SOE_up})-(\ref{eq:UL_SOE_down_LIM}) compute the confidence interval of the SOE forecast and ensure the feasible operation in the considered intervals.

Separating the charging and discharging components of the power offset allows for taking into consideration charging and discharging efficiency. Relaxation according to \cite{haessig_convex_2021}, visible in \cref{eq:UL_Objective} forces only one of the two decision variables to be different from zero, so that \cref{eq:UL_BESSPower} is compatible with the definition in \cref{eq:B_g}. Equations (\ref{eq:UL_limitCharging})-(\ref{eq:UL_limitDischarging}) ensure the final power of the battery to be within the operational limits.  Finally, \cref{eq:UL_BESSPower} computes the final power output of the BESS, positive if charging. The offset is directly applied to modify the dispatch plan of the HPP, as visible in \cref{fig:Framework}.
\subsection{Lower Layer MPC (LLMPC)}
\label{sec:LL}
The lower layer is a rolling-horizon MPC responsible for computing in real-time (i.e., each second) the set-point splitting policy between the HPP and the BESS, with the following objectives:
\begin{enumerate}[label=\alph*)]
\item Optimal tracking of the FCR provision;
\item minimize the number of movements and the total mileage of the hydropower servomechanisms;
\item ensure the BESS to operate within its physical limits.
\end{enumerate}
For every time-step:
\begin{equation}
    k \in [k,k+p] \nonumber,
\end{equation} where $p$ is the length of the MPC horizon, the expected power output $P_k^{\text{set}}$ of the hybrid system is:
\begin{equation}
    {P}^{\text{set}}_{k} = {P}^{\text{disp}}_{k} + \left[ 50 - \hat{f}_k \right] \cdot \sigma_f
\end{equation}
where the term term $\hat{f}_k$ indicates the expected value of the grid frequency at time $k$. Information about this short-term frequency forecast is contained in Appendix \ref{Appendix:FrequencyForecasting}. As a consequence, the tracking error $\text{TE}_k$ is the difference between the expected power output and the actual aggregated production of HPP + BESS, i.e. the power flow at the \emph{Point of Common Coupling} (PCC):
\begin{equation}
    \text{TE}_{k} = {P}^{\text{set}}_{k} - \big({H}_{k} - {B}_{k}\big)
    \label{eq:DispatchTracking_k}
\end{equation}
and, over the entire MPC window:
\begin{equation}
    \mathbf{TE}_{k:k+p} = \lVert \mathbf{P}^{\text{set}}_{k:k+p} - \big(\mathbf{H}_{k:k+p} - \mathbf{B}_{k:k+p}\big) \rVert_2
    \label{eq:DispatchTracking}
\end{equation}
The second objective, i.e. the reduction of the number of movements of the hydropower actuators, can be modeled as the minimization of the cardinality of the array containing the variation of the hydroelectric unit power output with respect to the previous time instant:
\begin{equation}
    \min \quad \text{card}(\mathbf{\Delta H}_{k:k+p})
    \label{eq:MovementReduction}
\end{equation}
with $\Delta H_k = H_k - H_{k-1} \quad \forall k \in [k,k+p]$. As discussed in \cite{boyd_convex_2004}, the cardinality of a quantity can be relaxed with its $\ell_1$-norm. In other words, $\lVert \mathbf{\Delta H}_{k:k+p} \rVert_1$ is the convex envelope of the objective function of \cref{eq:MovementReduction}. The latter equation can therefore be relaxed, and assumes the following form:
\begin{equation} 
   \min \quad  \gamma\lVert \mathbf{\Delta H}_{k:k+p} \rVert_1
     \label{eq:MovementReductionRelaxed}  
\end{equation}
where $\gamma$ is non-negative parameter tuned to achieve the desired sparsity. Equations (\ref{eq:DispatchTracking}) and (\ref{eq:MovementReductionRelaxed}) constitute the objective of the lower-layer MPC. 

The constraints of the OP ensure that the BESS and the HPP operate within their physical limits. In particular, the capability curve $\zeta$ of the BESS converter is considered as a function of the voltage on both the DC and AC side ($v_k^{DC}$ and $v_k^{AC}$, respectively) and of the BESS SOE:
\begin{equation}
    B_k \le \zeta(v_k^{DC},v_k^{AC},SOE_k) \label{eq:LL_pqcurve_eq}
\end{equation}Similarly to \cite{gerini_optimal_2022}, the estimation of the BESS $v_k^{DC}$ is based on the battery Three Time Constant (TTC) whose dynamic evolution can be expressed as a linear function of battery current by applying the transition matrices $\phi^v$,$\psi^v$,$\psi_1^v$ (see \cref{eq:LL_estVdc}). In the TCC model, the quantity $x_k$ is the state vector of the voltage model \footnote{This modelling choice of modeling, based on \cite{sossan_achieving_2016}, offers greater accuracy than the commonly used two-time constant model \cite{bahramipanah_enhanced_2014}. However, for BESS predictions with a 1-second horizon, simpler models can be utilized, as discussed in Section \ref{sec:Experimental Validation}.}, and the DC current is expressed as $i_k^{DC}$. The latter is computed in \cref{eq:LL_estIdc}. The active power at the DC bus $B_k^{DC}$ is related to the active power set-point AC side of the converter $B_k^{\text{set}}$ according to \eqref{eq:LL_Pac2Pdc}. The magnitude of the direct sequence component $v_k^{AC}$ of the phase-to-phase voltage on the AC side of the converter, is assumed to be equal to the last available measurement, as indicated in \eqref{eq:LL_EstVac}. Finally, the BESS SOE evolution is computed in \cref{eq:LL_SOEssm} in ensured within its physical limits by Eqs. (\ref{eq:LL_SOE_lim}).

For the HPP, a first-order discrete-time dynamical system is used to model the response $H_k$ to a set point $H_k^{\text{set}}$ considering a time constant $\tau_H$:
\begin{equation}
    H_{k}  = (1 - \frac{\Delta k}{\tau_H}) \cdot H_{k-1} + \frac{\Delta k}{\tau_H} \cdot H_{k}^{\text{set}} \label{eq:LL_1OrderResponseHydroeq}
\end{equation} where $\Delta k$ is the time interval between two consecutive discrete-time samples (i.e., one second). The HPP output power is limited within $H_{\min}$ and $H_{\max}$ by \cref{eq:LL_HydroPower_lim} while \cref{eq:LL_HydroRate_Lim} ensures the power ramping rate to be lower than its maximum allowed value $\dot{H}_{\max}$, expressed in \SI{}{\kilo\watt\per\second}. As we assume the dispatch plan to include concession head control and given the short-term horizon of the MPC, no constraints on the concession head are introduced\footnote{In the case of need for modelling constraints on the concession head, similarly to what done in \cite{borghetti_maximum_2008}, information about the hill-chart of the runner are needed, to translate power set-points into discharge values.}. The LLMPC, running every second $k$, is the following OP:
\begin{subequations}
\begin{alignat}{1}
    \big[\mathbf{H}^{\text{set}}_{k:k+p}, &\mathbf{B}^{\text{set}}_{k:k+p} \big]  =  \nonumber\\
    &= \underset{\mathbf{H}, \mathbf{B} \in \mathbb{R}^p}{\arg \min } \quad \mathbf{TE}_{k:k+p} + \gamma\cdot\lVert \mathbf{\Delta H}_{k:k+p} \rVert_1\\
\text { subject to:} & \nonumber\\
B_k & \le \zeta(v_k^{DC},v_k^{AC},SOE_k) \label{eq:LL_pqcurve}\\
B_k^{DC} & = \left\{\begin{array}{cc}
         B_k^{\text{set}}\cdot\eta_{\text{ch}}, & \forall B_k^{\text{set}} \ge 0  \\
         B_k^{\text{set}}/\eta_{\text{dch}} , & \forall B_k^{\text{set}} < 0
    \end{array}  \right.  \label{eq:LL_Pac2Pdc} \\
v_{k}^{DC}& = \phi_v x_{k-1}    +\psi_i^v{i}_k^{DC}+\psi_1^v\bm{1} \label{eq:LL_estVdc}\\
{i}_k^{DC} & \approx \frac{{B}_k^{DC}}{v_{k}^{DC} } \label{eq:LL_estIdc} \\
v_k^{AC} & \approx v_{k-1}^{AC} \label{eq:LL_EstVac} \\ 
SOE_{k} & = SOE_{k-1} +i^{DC}_k v_{k}^{DC}  \frac{1}{C_{B}} \label{eq:LL_SOEssm}\\
SOE_{\min} &  \leq SOE_k \leq  SOE_{\max} \label{eq:LL_SOE_lim}\\
H_{k} & = (1 - \frac{\Delta k}{\tau_H}) \cdot H_{k-1} + \frac{\Delta k}{\tau_H} \cdot H_{k}^{\text{set}} \label{eq:LL_1OrderResponseHydro}
\\
H_{\min} & \leq H_k \leq  H_{\max} \label{eq:LL_HydroPower_lim}\\
-\dot H_{\max} &\leq \Delta H_k \leq \dot H_{\max}\label{eq:LL_HydroRate_Lim} \\
      & \hspace{1.6cm}\forall k  \in [k,\ldots,k+p] \nonumber
    \end{alignat}
    \label{eq:LowerLayerMPC}
\end{subequations}
A way to convexify constraints (\ref{eq:LL_pqcurve}) - (\ref{eq:LL_EstVac}) has been presented in \cite{zecchino_optimal_2021} and used in \cite{gerini_optimal_2022}. The optimization problem is solved at each time step $k$ (with updated information) on a sliding horizon from the index $k$ to $k+p$. At each $k$, the control trajectories for HPP and BESS for the whole residual horizon $\big[\mathbf{H}^{\text{set}}_{k:k+p}, \;\mathbf{B}^{\text{set}}_{k:k+p} \big]$ are available. However, only the first components, denoted by:$ \big[{H}^{\text{set}}_{k}, {B}^{\text{set}}_{k} \big] $ are considered for actuation.

\section{Experimental Validation}
\label{sec:Experimental Validation}
\subsection{Experimental setup}
\label{sec:Experimental_setup}
\begin{figure}[ht]
\centering
\includegraphics[width=0.9\linewidth]{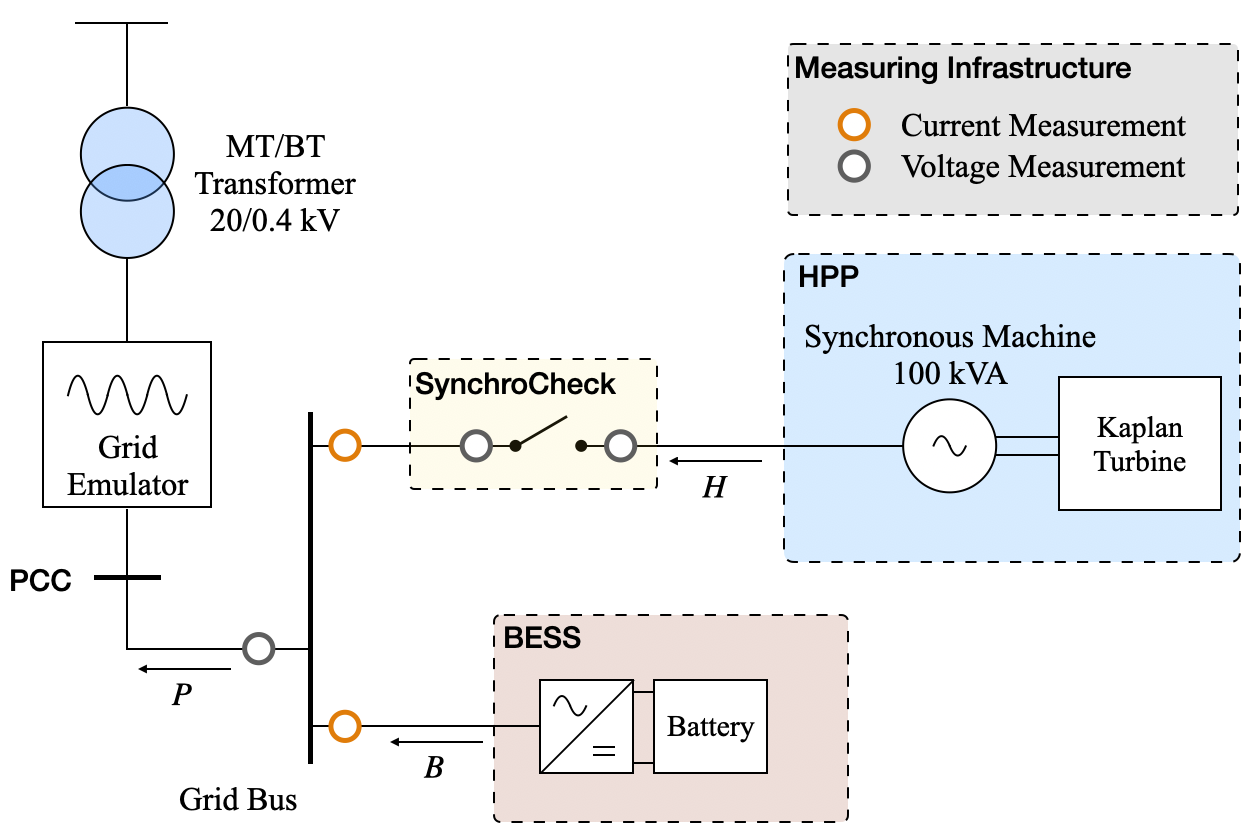}
\caption{Schematics of the PTMH PF3 power grid used to carry out the experiments}
\label{fig:Schematics}
\end{figure}

The hybrid experimental platform used modernizes an existing platform in the \emph{Plateforme Technologique Machines Hydrauliques} (PTMH), at EPFL. The current platform is a closed-loop test rig that allows performance assessments of hydraulic machines in the four-quadrant characteristic curve with an accuracy of 0.2 \%, complying with the IEC60153 standard for the testing of reduced scale physical models. The specific hydraulic energy in the closed-loop test-rig is generated by two centrifugal pumps. They allow for a maximum head of 100 m and a maximum discharge of 1.4 m$^3$s$^{-1}$. Furthermore, the pressure in the draft tube is set by adjusting the pressure in the downstream reservoir by using a vacuum pump. The hydroelectric unit governing systems is built as a standard speed governor \cite{grigsby_power_2007} with frequency droop $\sigma_f =  \SI{125}{\kilo\watt\per\hertz}$ and a dead band of $\SI{2}{\milli\hertz}$. The choice of droop is dictated by the similarity need with the prototype of the reduced scale model turbine, installed in Vogelgrun (FR) \cite{kadam_hybridization_2023}, and comprehensively discussed in the Appendix \ref{Appendix:BESS_Sizing}. The measurement infrastructure employs a distributed sensing system based on \emph{Phasor Measurement Units} (PMUs). This system allows for real-time acquisition of precise power flow data thanks to the PMUs' high reporting rate of 50 frames per second and remarkable accuracy, with a standard deviation equivalent to $0.001$ degrees (approximately 18 $\mu$rad) and error in the frequency estimation $\le \SI{0.4}{\milli\hertz}$  \cite{romano_enhanced_2014}. A 50 kW Kaplan turbine is connected to the grid bus with a synchronous machine, rated 100 kVA. On the same bus, the BESS is connected. For more information, see \cite{gerini_experimental_2021}. The frequency of the grid is regulated by a grid emulator with a nominal power rating of 100 kVA. The configuration of the hybrid system is illustrated in Figure \ref{fig:Schematics}. As the turbine is connected to the grid through a synchronous machine, a synchro-check mechanism is necessary. The comprehensive list of measurements is detailed in Table \ref{tab:measurements}.


\begin{table*}[ht]
    \centering
    \caption{Measurements on the PTMH PF3}
    \begin{tabular}{l c l c l c}
    \toprule
    \multicolumn{2}{c}{\textbf{Hydropower}} & \multicolumn{2}{c}{\textbf{BESS}} & \multicolumn{2}{c}{\textbf{PMU}} \\
    \cmidrule(lr){1-2} \cmidrule(lr){3-4} \cmidrule(lr){5-6}
    \textbf{Name} & \textbf{Measurement Unit} & \textbf{Name} & \textbf{Measurement Unit} & \textbf{Name} & \textbf{Measurement Unit} \\
    \midrule
    Discharge & \SI{}{\cubic\meter\per\second} & State of Charge & \SI{}{\percent} & Voltage Magnitude & \SI{}{\volt}\\
    Turbine Speed & \SI{}{\rpm} & DC Current & \SI{}{\ampere} & Voltage Angle & \SI{}{\radian} \\
    Head & \SI{}{\meter} & DC Power & \SI{}{\watt} & Current Magnitude & \SI{}{\ampere} \\
    Hydraulic Efficiency & \SI{}{\percent} & DC Voltage & \SI{}{\volt} & Current Angle & \SI{}{\radian} \\
    Hydraulic Power & \SI{}{\watt} & Active power & \SI{}{\watt} & Active Power & \SI{}{\watt} \\
    Shaft Torque & \SI{}{\newton\meter} & Reactive Power & \SI{}{\var} & Reactive Power & \SI{}{\var} \\
    Mechanical Power & \SI{}{\watt} & Temperature & \SI{}{\degreeCelsius} & Frequency & \SI{}{\hertz} \\
    Guide Vanes Opening (GVO) & \SI{}{\deg} & & & & \\
    Runner Blade Angle (RBA) & \SI{}{\deg} & & & & \\
    Runner Blade Torque (RBT) & \SI{}{\newton\meter} & & & & \\
    \bottomrule
    \end{tabular}
    \label{tab:measurements}
\end{table*}

\subsection{Experimental Tests}
\label{sec:Experimental_tests}
\begin{figure*}
\centering
\includegraphics[width=0.9\linewidth]{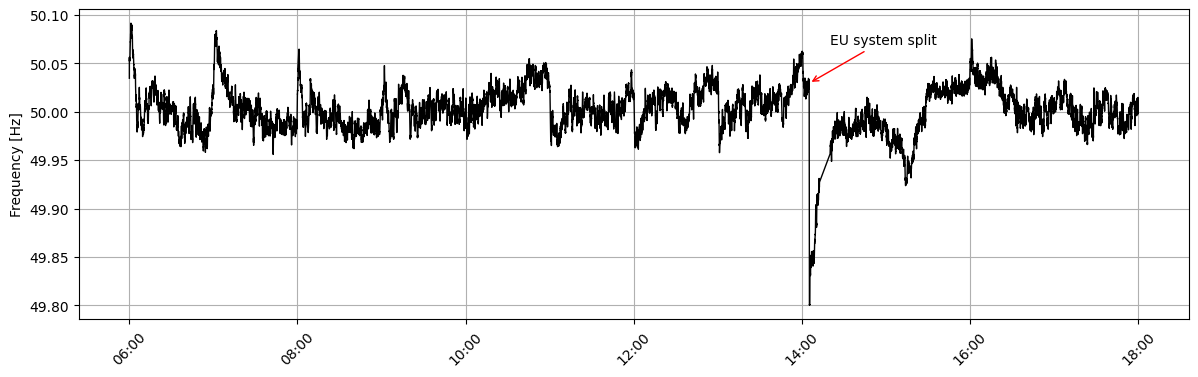}
\caption{Frequency Time Series for the test. Measurement from SwissGrid during the system split of the 8th of January 2021 \cite{ics_investigation_expert_panel_continental_2021}}
\label{fig:frequency}
\end{figure*}

In this section, we present the experimental campaign to validate the proposed framework. A series of 12 hour-long tests are performed under the same grid condition. The frequency time-series enforced at the PCC by the grid emulator is illustrated in \cref{fig:frequency}. This frequency data corresponds to measurements taken on January 8th, 2021, when the ENTSO continental Europe synchronous area experienced a system split. This particular time frame is chosen to ensure the inclusion of typical daily frequency patterns, for the first 8h of test, and also to subject the system to more challenging scenarios in the remaining 4 hours. More information about the system split event can be found in \cite{ics_investigation_expert_panel_continental_2021}.
Under the described grid conditions, the following tests are conducted:
\begin{enumerate}[label=\alph*)]
    \item Kaplan unit operating alone.
    \item Hybrid Kaplan unit with a \SI{5}{\kilo\watt}/\SI{5}{\kilo\watt\hour} BESS, controlled by a \emph{Dead Band Filter} (DBF).
    \item Hybrid Kaplan unit with a \SI{9}{\kilo\watt}/\SI{9}{\kilo\watt\hour} BESS, controlled by DBF.
    \item Hybrid Kaplan unit with a \SI{5}{\kilo\watt}/\SI{5}{\kilo\watt\hour} BESS, controlled by DLMPC.
    \item Hybrid Kaplan unit with a \SI{9}{\kilo\watt}/\SI{9}{\kilo\watt\hour} BESS, controlled by DLMPC.
\end{enumerate}where the DBF control approach consists in the use of a dead-band filter on the frequency signal input of the governor \cite{kadam_hybridization_2023}. Under this scheme, any frequency deviation lower than the one associated with the maximum power capacity of the BESS is allocated to the BESS itself, while any deviation exceeding this threshold is directed to the hydroelectric unit. For instance, in the context of a \SI{5}{\kilo\watt} BESS, the frequency threshold for the DBF control is at \SI{40}{\milli\hertz}, and for a \SI{9}{\kilo\watt} BESS, it is set at \SI{72}{\milli\hertz}. For more details on the sizing of the BESS and the determination of these threshold values, please refer to Appendix \ref{Appendix:BESS_Sizing}. Even when utilizing the DBF control strategy, an upper layer is implemented to function as a SOE manager of the BESS. All the tests are performed considering a flat dispatch plan of \SI{27}{\kilo\watt}, constant for the 12 hours, to ensure that the wear and tear analysis of the movements is only related to FCR support. Moreover, the head of the system is kept constant at $\SI{10}{\meter}$.

As for the DLMPC approach, the ULMPC is executed hourly based on the energy prediction for regulation, as outlined in Appendix \ref{Appendix:FrequencyForecasting}. On the other hand, the LLMPC runs every second using a sliding window spanning 30 seconds. Additional information about short-term frequency forecasting is also detailed in Appendix \ref{Appendix:FrequencyForecasting}.
\subsection{Results}
This analysis aims to highlight the advantages of BESS hybridization and to conduct a comparative assessment of distinct control techniques across various \emph{Key Performance Indicators} (KPIs). This comprehensive analysis covers three primary aspects: the quality of the FCR provision, the mitigation of wear and tear on the hydroelectric governing system, and the safe operation of the BESS.
\subsubsection{FCR provision quality}
To assess the effectiveness of the FCR provision, we introduce the \emph{Root Mean Squared} (RMS) of the tracking error $\text{TE}$, as defined in \cref{eq:DispatchTracking_k}, computed over the full experiment time horizon. 
By integrating over a certain amount of time $\Delta t$ the power error it is possible to estimate a mean energy error $E_{k}^{k+\Delta t}$, as follows:
\begin{equation}
    E_{k}^{k+\Delta t} = \frac{1}{\Delta t} \sum_{k}^{k+\Delta t}\text{TE}_k 
\end{equation}
The resulting power error and energy error values, computed for $ \Delta t = [10, 30 ,60]$ seconds are presented in Table \ref{tab:rmse-values-reduction}. While the RMS of $\text{TE}$ might be influenced by power measurement noise, the energy errors consistently underline the benefits of hybridization in enhancing FCR provision quality.
\begin{table*}[t]
    \centering
    \caption{RMSE Values and Reduction (\%) with Respect to Only Hydro}
    \begin{tabular}{l c | c c c c}
    \toprule
    \textbf{Configuration} & \textbf{Power Error} & \multicolumn{3}{c}{\textbf{Energy Error }}\\
     & (RMS) & \multicolumn{3}{c}{(RMS)}\\
    \textbf{} & \textbf{1s} & \textbf{10s} & \textbf{30s} & \textbf{60s} \\
    \midrule
    \textcolor{blue}{DBF (\SI{5}{\kilo\watt})} & 0.2193 ($-8.91\%$) & 0.3447 ($-43.07\%$) & 0.3872 ($-47.01\%$) & 0.4095 ($-47.48\%$) \\
    \textcolor{orange}{DBF (\SI{9}{\kilo\watt})} & 0.2091 ($-13.02\%$) & 0.3032 ($-49.76\%$) & 0.3326 ($-54.61\%$) & 0.3544 ($-54.54\%$) \\
    \textcolor{green}{DLMPC (\SI{5}{\kilo\watt})} & 0.2348 ($-2.42\%$) & 0.2990 ($-50.56\%$) & 0.3696 ($-49.62\%$) & 0.3952 ($-49.30\%$) \\
    \textcolor{red}{DLMPC (\SI{9}{\kilo\watt})} & 0.2243 ($-6.59\%$) & 0.3182 ($-47.35\%$) & 0.3468 ($-52.65\%$) & 0.3089 ($-60.39\%$) \\
    \textcolor{purple}{Only Hydro} & 0.2405 ($+0.00\%$) & 0.6048 ($+0.00\%$) & 0.7331 ($+0.00\%$) & 0.7795 ($+0.00\%$) \\
    \bottomrule
    \end{tabular}
    \label{tab:rmse-values-reduction}
\end{table*}
Although the BESS size contributes to reducing the RMS of $\text{TE}$, the impact is not notably significant. Interestingly, DBF and DLMPC techniques can be considered equivalent in terms of set-point tracking quality. Nonetheless, for data aggregated at 10, 30, and 60-second intervals, the tracking error of any hybrid system demonstrates a reduction of at least 50\% in comparison to the error exhibited by the non-hybrid system.

\subsubsection{Wear and Tear Reduction}
\begin{figure}[ht]
    \centering
    \includegraphics[width=0.9\linewidth]{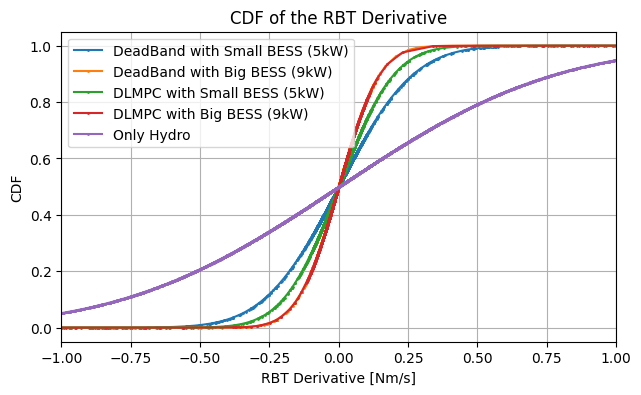}
    \caption{Comparative analysis of Blade Torque Oscillation for the different scenarios.}
    \label{fig:RBT}
\end{figure}

The assessment of wear reduction benefits is undertaken through the consideration of three specific KPIs: servomotors mileage, \emph{Number of Movements} (NoM) and torque oscillation on the blades. The reduction in both guide vane and runner blade NoM and mileage are significant indicators of the wear reduction achieved through hybridization \cite{valentin_benefits_2022}. Table \ref{tab:mileage-movements} provides a comprehensive overview of this wear reduction, underscoring the substantial advantages offered by the MPC technique. All the percentage values are computed relative to the baseline configuration, represented by 'Only Hydro'. Notably, the DLMPC control method demonstrates superior performance compared to the DBF control. The degree of improvement achieved through more sophisticated control techniques is particularly pronounced for smaller BESS sizes. In essence, when the BESS size is sufficiently large, the necessity for optimal control diminishes.  For instance, in the \SI{5}{\kilo\watt} BESS experiments, the DBF control attains a substantial reduction of 90.7\% in runner blade mileage and 91.6\% in the number of movements. Similarly, with the same BESS size, the DLMPC algorithm demonstrates even greater reductions, achieving 94.0\% and 97.1\% reduction in runner blade angle mileage and the number of movements, respectively. Comparable outcomes are observed when examining the corresponding reduction in guide vane servomotors. These findings highlight the benefits of implementing hybridization with advanced control techniques such as DLMPC, particularly standing out for its wear reduction performance for smaller BESS sizes. 
\begin{table*}[t]
    \centering
    \caption{Reduction in Mileage and Movements (\%)}
    \begin{tabular}{lcccc}
    \toprule
    \textbf{Configuration} & \multicolumn{2}{c}{\textbf{Guide Vanes Opening}} & \multicolumn{2}{c}{\textbf{Runner Blade Angle}} \\
     & \multicolumn{2}{c}{(GVO)} & \multicolumn{2}{c}{(RBA)} \\
    \cmidrule(lr){2-5}
      & \textbf{Mileage } & \textbf{Number of Movements} & \textbf{Mileage } & \textbf{Number of Movements} \\
    \midrule
    \textcolor{blue}{DBF (\SI{5}{\kilo\watt} BESS)} & 4.18 (-91.5\%) & 662 (-92.9\%) & 3.06 (-90.7\%) & 753 (-91.6\%) \\
    \textcolor{orange}{DBF (\SI{9}{\kilo\watt} BESS)} & 0.98 (-98.0\%) & 126 (-98.6\%) & 0.70 (-97.9\%) & 128 (-98.6\%) \\
    \textcolor{green}{DLMPC (\SI{5}{\kilo\watt} BESS)} & 3.33 (-93.2\%) & 292 (-96.8\%) & 1.98 (-94.0\%) & 258 (-97.1\%) \\
    \textcolor{red}{DLMPC (\SI{9}{\kilo\watt} BESS)} & 0.99 (-98.0\%) & 64 (-99.3\%) & 0.61 (-98.2\%) & 52 (-99.4\%) \\
    \textcolor{purple}{Only Hydro} & 49.03 (+0.00\%) & 9261 (+0.00\%) & 32.93 (+0.00\%) & 8970 (+0.00\%) \\
    \bottomrule
    \end{tabular}
    \label{tab:mileage-movements}
\end{table*}
Finally, for an estimate of the forces that affect the blades and their servomechanisms, the torque oscillations occurring on the runner blades are evaluated. To capture torque data, strain gauges are employed on the runner blades. As outlined in \cite{presas_use_2021}, strain gauge testing emerges as the sole reliable approach for acquiring both static and dynamic stress information from the runner. Over recent years, this method has gained widespread acceptance as a standard practice for newly installed runners, as highlighted in \cite{presas_fatigue_2019}. The \emph{Cumulative Density Function} (CDF) of the blade torque derivative is analyzed in \cref{fig:RBT}. In line with the trends observed in other wear reduction KPIs, the sole hydro case demonstrated inferior performance, characterized by elevated levels of torque oscillations on the blades. In the case of the 5 kW BESS, it's noticeable that the CDF of the blade torque derivative is significantly narrower under the DLMPC control strategy compared to DBF. This narrower distribution implies that there are fewer occurrences of high torque oscillations, indicating improved performance in mitigating such oscillations with DLMPC. The presence of a larger BESS size demonstrates a positive impact on the torque oscillation reduction.
\subsubsection{Safe BESS Operation}
\begin{figure*}[ht]
    \centering
    \includegraphics[width=0.87\linewidth]{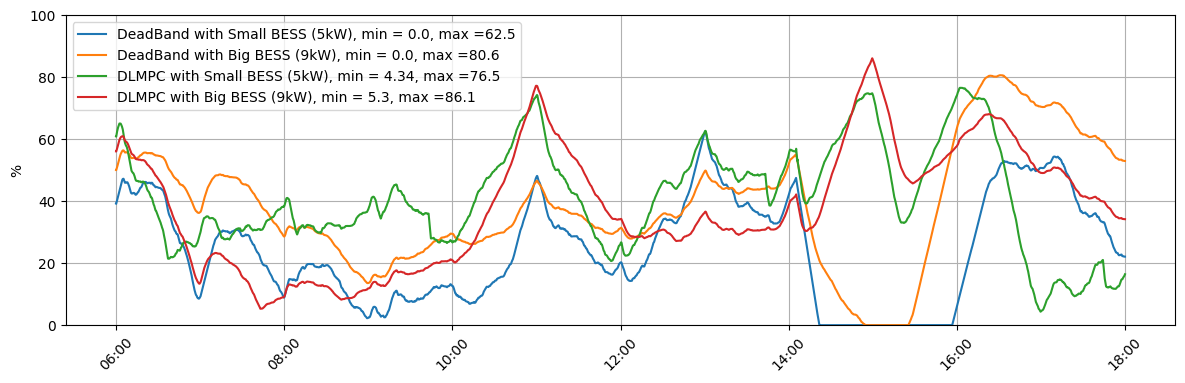}
    \caption{BESS SOE evolution during the different experiments}
    \label{fig:SOE}
\end{figure*}

The BESS SOE evolution over the 12 hours, for each experiment, is illustrated in \cref{fig:SOE}. It is worth noting that, the two DBF experiments operate effectively until the grid split event occurs, around 14h00. Indeed, the regulating energy prediction from \cref{eq:Wf}, is designed to function correctly in approximately 95\% of cases (see \cref{tab:SARIMA}) under normal grid conditions. However, in the event of a grid disruption, the power grid dynamics drastically change, and the new grid configuration differs from the one on which the SARIMA model was originally trained. Consequently, both DBF control strategies are unable to control the SOE within its limits during the latter half of the experiment. In contrast, the DLMPC strategy continually monitors the BESS SOE by means of the LLMPC. This frequent monitoring guarantees that the BESS operates within its predefined operational limits, even if the FCR enrgy prediction $E_{\text{FCR}}$ deviates from the expected values. This feature substantially bolsters the system's reliability, allowing it to adapt to unforeseen grid dynamics.

\section{Conclusion}
\label{sec:Conclusion}
In this paper, we present a comprehensive solution to address operational challenges in Run-of-River Hydropower Plants (RoR HPPs) related to continuous power regulations due to Frequency Containment Reserve (FCR) provision. By integrating Battery Energy Storage Systems (BESS) with RoR HPPs using a double-layer Model Predictive Control (MPC) approach, we validate a novel strategy to enhance FCR capabilities and simultaneously reduce the wear and tear. The proposed control framework consists of an upper layer MPC responsible for State-of-Charge management of the BESS and a lower layer MPC for optimal splitting policy of power set points between the turbine and the BESS. Rigorous reduced-scale experiments conducted in an innovative testing platform validated the effectiveness of the proposed MPC-based control strategy across diverse grid scenarios and BESS sizing. The experimental results showcased the superiority of the double-layer MPC strategy over simpler techniques such as dead-band control. The hybridized system showcased enhanced FCR provision quality and a notable reduction in servomechanism stress. The analysis of wear reduction revealed a substantial decrease in both guide vane and runner blade angle mileage, ranging from 93.2\% with a 5 kW BESS to as much as 98\% for a 9 kW BESS. Similar reductions were observed in the case of movements. Moreover, reduced torque oscillations on the blades further emphasize the benefits of the proposed hybridization approach. Future work on the subject involves an experimental comparison between the hybridization scenario outlined and the implementation of variable speed for the Kaplan turbine used as a propeller, i.e. with fixed blades.


\ifCLASSOPTIONcaptionsoff
  \newpage
\fi

\bibliographystyle{IEEEtran}
\bibliography{bib.bib}
\appendices
\section{Frequency Forecasting}
\label{Appendix:FrequencyForecasting}
\subsection{Regulating Energy Forecasting}
\begin{figure}[ht]
\centering
\includegraphics[width=0.86\linewidth]{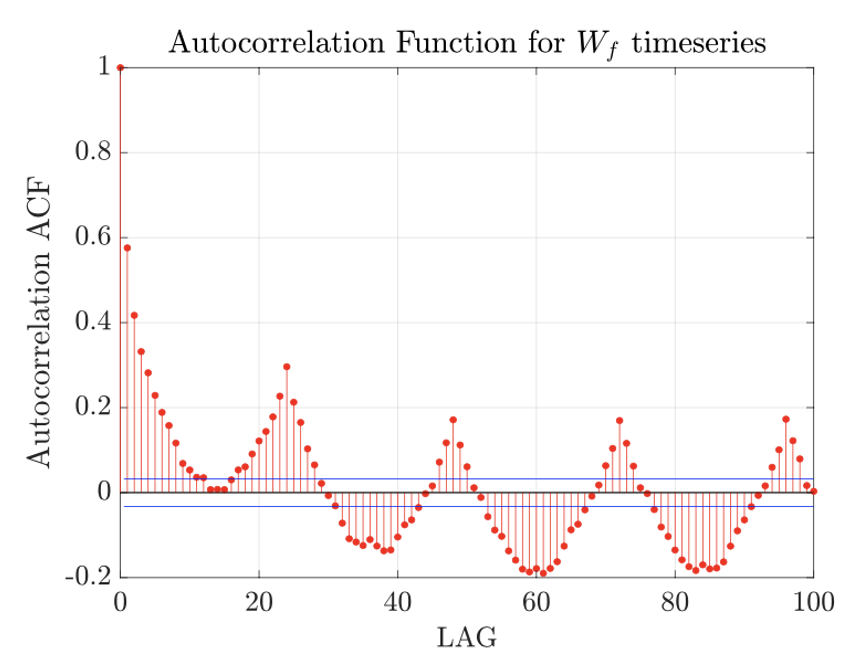}
\vspace{-0.3cm}
\includegraphics[width=0.86\linewidth]{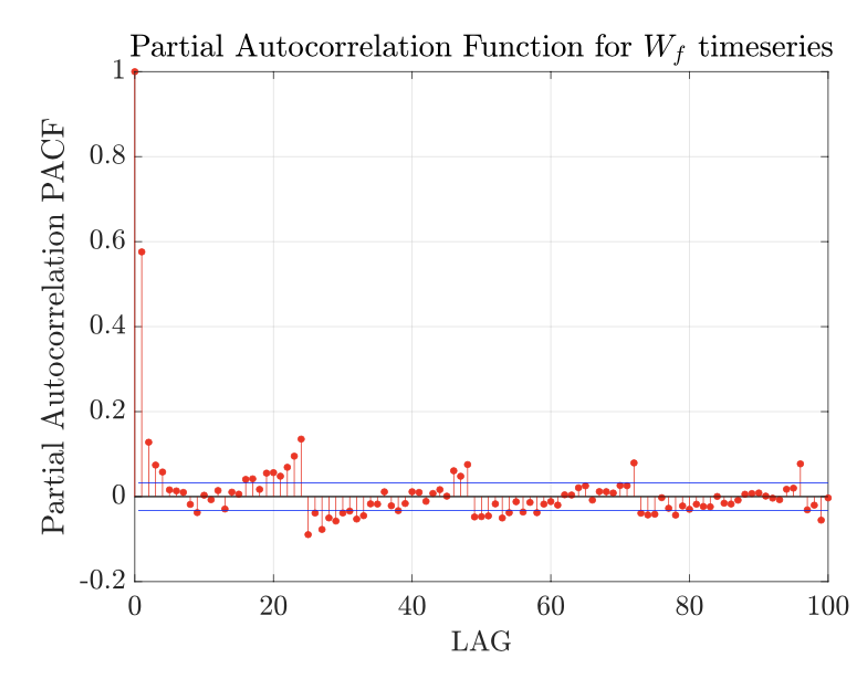}
\caption{$W_g$ Auto-Correlation Function (ACF) above and Partial Auto-Correlation Function (PACF) below.}
\label{fig:AC}
\end{figure}

Both the upper-layer and lower-layer MPC rely on distinct types of frequency forecasting. In the upper layer MPC, a forecast of the FCR regulating energy for the next hour is necessary. This entails estimating the integral of the frequency deviation over time, once the droop is defined. The ability to predict this quantity was initially introduced in \cite{piero_schiapparelli_quantification_2018} and subsequently adopted by various researchers, including \cite{gerini_improving_2021}. In this contribution, we propose an enhanced model for frequency forecasting and compare its performance with the model from \cite{piero_schiapparelli_quantification_2018} in terms of standard deviation. The formulation for FCR energy estimation during an hour $g$ is given by:
\begin{equation}
E_{\text{FCR}} = \int_g P_{\text{FCR}} dt = \int_g \sigma_f \cdot \Delta f dt = \sigma_f W_g,
\label{eq:Wf}
\end{equation}
where $\sigma_f$ is the frequency droop. The motivation for updating the models stems from a more comprehensive statistical analysis carried out by the authors using a one-year-long set of frequency time series (from March 2019 to April 2020). This analysis clearly demonstrates a discernible seasonality effect on a daily basis, as evident from \cref{fig:AC}, which was not considered in the model of \cite{piero_schiapparelli_quantification_2018}.
For this reason, a new Seasonal Auto Regressive Integrated Moving Average (SARIMA) model is introduced. We present the model and its performance over a testing set of data (from September 2020 to December 2020) in \cref{tab:SARIMA}. In this table, $\gamma_T$ indicates the time intervals at which the residuals exceed the corresponding 95\% or 99\% confidence intervals. It is defined as follows:
\begin{equation}
\gamma_T = \frac{\text{Time with } \lvert {r_i}\rvert \ge k\sigma}{\text{Total observation time}},
\end{equation}
where $\sigma$ represents the variance and $k$ takes values 2 and 3 for the 95\% and 99\% intervals, respectively.

\setlength{\tabcolsep}{10pt}
\renewcommand{\arraystretch}{1.5}
\begin{table}[ht]
    \centering
   \begin{tabular}{|c|c|c|cc|}
\hline Models&$\sigma$&$\rho$&$\gamma_T$&MSE\\
\hline \multirow{2}{*}{ARIMA$(8,0,0)(0,0,0)$} & 25 & $95 \%$ & $7.2 \%$ & 562 \\
& 25 & $99 \%$ & $1.5 \%$ & 562 \\
\hline \multirow{2}{*}{ SARIMA$(6,0,0)(0,1,1)_{24}$} & 22 & $95 \%$ & $7.2 \%$ & 485 \\
& 22 & $99 \%$ & $2.1 \%$ & 485 \\
\hline \multicolumn{5}{|c|}{ Dataset: PMU-5, EPFL, September 2020 - December 2020 \cite{zanni_pmu-based_2020}} \\
\hline
\end{tabular}
\vspace{0.1cm}
    \caption{Comparison between the forecasting model used in the upper layer MPC and \cite{piero_schiapparelli_quantification_2018}.}
    \label{tab:SARIMA}
\end{table}
Lastly, a residual analysis of the SARIMA model is conducted, to demonstrate that the proposed model cannot be further improved by increasing the order of the model. The residuals exhibit a zero-mean normal distribution. Furthermore, a Durbin-Watson test confirms the absence of correlation among the residuals. The test statistic is computed as:
\begin{equation}
s = \frac{\sum_{i=2}^N\left(r_i-r_{i-1}\right)^2}{\sum_{i=1}^N\left(r_i\right)^2} = 2.0051.
\nonumber
\end{equation}
Given its enhanced performance, the forecaster selected to produce the regulating energy prediction fed to the upper layer MPC during the experimental validation is $\text{SARIMA}(6,0,0)(0,1,1)_{24}$.
\subsection{Short-Term Frequency Forecasting}
Frequency prediction is a crucial component of the lower-layer MPC. Specifically, the experimental validation that employs the lower layer MPC necessitates second-interval predictions for a 30-second horizon. The inherent stochastic behavior of grid frequency in bulk power systems poses a significant challenge in achieving accurate predictions. In fact, only a limited number of studies have explored the feasibility of such predictions. One exception can be found in \cite{chettibi_real-time_2021}, which introduces the real-time prediction of grid voltage and frequency using artificial neural networks. This method primarily focuses on short-term frequency prediction (0.183 ms and 1 sec), demonstrating satisfactory \emph{Root Mean Squared Error} (RMSE) values for both one-step and three-step ahead predictions. However, the RMSE achieved through this approach is on par with the RMSE obtained from a simple AR(0) regression model. This regression model essentially assumes continuity, implying that the subsequent frequency measurement remains consistent with the preceding one. In details, \cite{chettibi_real-time_2021} claims an RMSE$= 0.0039$ Hz of their one-sec ahead forecaster tested for during one day of May 2019. AR(0) regression model for secondly-sampled frequency timeseries for the same month (data from: \cite{ngeso_historic_2019}) results in an RMSE $= 0.0024$. However, given that the data-set utilized in \cite{chettibi_real-time_2021} is not publicly accessible, offering more comprehensive insights presents a challenge.
Consequently, the frequency predictor for the lower-layer MPC does not hinge on genuine forecasting, but rather assumes that the forthcoming 30 seconds of frequency will mirror the preceding values. 

\section{BESS Sizing for Experimental Validation}
\label{Appendix:BESS_Sizing}

Optimal BESS sizing for this application is beyond the scope of this study, as it requires comprehensive consideration of various factors and extended time frames. However, the authors wish to clarify the rationale behind the BESS sizing during for the experimental campaign. As delineated in \cite{valentin_hybridization_2022}, the hydro-turbine governor at Vogelgrun Hydropower Plant (HPP) is characterized by a regulating energy of \SI{3.5}{\mega\watt} (10\% of its nominal power) for a \SI{200}{\milli\hertz} frequency deviation, i.e. a droop of \SI{17.5}{\mega\watt\per\hertz}. Scaling this to the tested reduced-scale model yields a \SI{25}{\kilo\watt\per\hertz} droop with \SI{5}{\kilo\watt} of regulating power at \SI{200}{\milli\hertz} deviation. However, this regulating power isn't feasible due to battery size constraints and potential interference with power oscillation noise.

\begin{figure}[!h]
\centering
\includegraphics[width=0.95\linewidth]{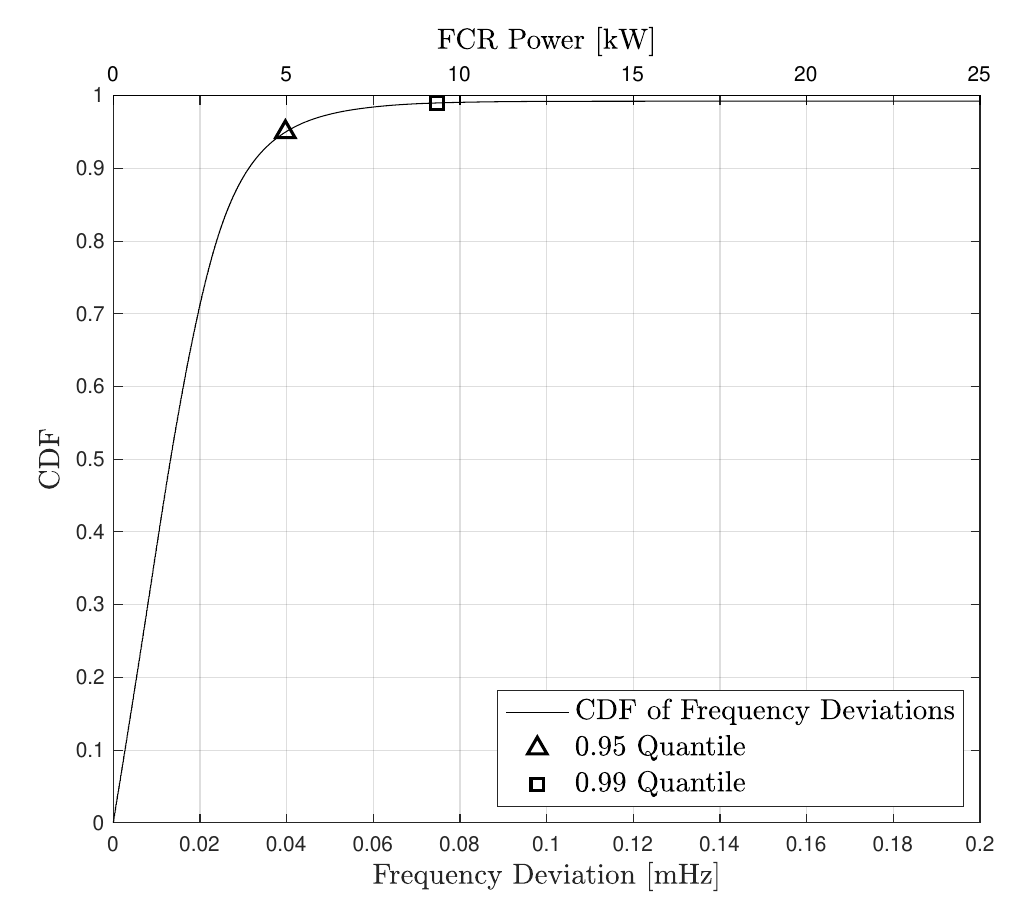}
\caption{CDF of Frequency deviation in continental Europe, based on year-long time series collected between 2019 and 2020. On the upper x-axis the FCR power associated with the frequency deviation.}
\label{fig:freuquencyCDF}
\end{figure}

To address the latter probelm, the facility's droop is increased by a factor of 5, leading to a \SI{125}{\kilo\watt\per\hertz} final droop. In accordance with \cite{valentin_hybridization_2022}, the installed BESS at Vogelgrun constitutes 16.5\% of the regulating energy, equivalent to \SI{4.125}{\kilo\watt}. For experimentation, we opted for two BESS sizes: \SI{5}{\kilo\watt} and \SI{9}{\kilo\watt}. The BESS power rating is the consequence of a statistical analysis of frequency time-series in the continental Europe power system, aiming to cover 95\% and 99\% of frequency deviations. Figure \ref{fig:freuquencyCDF} offers a visual representation of the CDF of frequency deviation in the ENTSO synchronous area, together with the power ratings derived by assuming a droop of \SI{125}{\kilo\watt\per\hertz}. Regarding the BESS energy sizing, it is important to note that this assumption impacts the outcomes of the DLMPC. Specifically, a BESS with more energy capacity would necessitate fewer interventions from the DLMPC, whereas a BESS with lower energy capacity would require more frequent actions. Due to the complexity of the tests, we did not conduct experiments with varying energy ratings for each BESS power rating. Instead, we made the assumption of a power-to-energy ratio of 1, aligning with market availability as mentioned in \cite{hesse_lithium-ion_2017}. Such an assumption is possible since the scope of the paper is not to provide optimal sizing, but rather to validate the control framework. Future studies will focus on evaluating the optimal BESS sizing for addressing challenges related to BESS hybridization in HPPs.

\end{document}